\documentclass[11pt]{book}

\usepackage{amssymb}
\usepackage{amsmath}
\usepackage{amsfonts}
\usepackage{epsfig}
\makeatletter

\renewenvironment{thebibliography}[1]
     {\section*{\bibname}%
      \@mkboth{\MakeUppercase\bibname}{\MakeUppercase\bibname}%
      \list{\@biblabel{\@arabic\c@enumiv}}%
           {\settowidth\labelwidth{\@biblabel{#1}}%
            \leftmargin\labelwidth
            \advance\leftmargin\labelsep
            \@openbib@code
            \usecounter{enumiv}%
            \let\p@enumiv\@empty
            \renewcommand\theenumiv{\@arabic\c@enumiv}}%
      \sloppy
      \clubpenalty4000
      \@clubpenalty \clubpenalty
      \widowpenalty4000%
      \sfcode`\.\@m}
     {\def\@noitemerr
       {\@latex@warning{Empty `thebibliography' environment}}%
      \endlist}

\makeatother

\newcommand{\sect}[1]{\setcounter{equation}{0}\section{#1}}

\renewcommand\bibname{References}

\begin{document}

\setlength{\baselineskip}{5.0mm}


\chapter[ ]{ Random matrix representations of critical statistics.}
\thispagestyle{empty}

\ \\

\noindent {{\sc V.E.Kravtsov}$^1$
\\~\\$^1$Abdus Salam International Center for Theoretical Physics,
\newline
P.O.Box586, 34100 Trieste, Italy }

\begin{center}
{\bf Abstract}
\end{center}
We consider two random matrix ensembles which are relevant for
describing critical spectral statistics in systems with multifractal
eigenfunction statistics. One of them is the Gaussian non-invariant
ensemble which eigenfunction statistics is multifractal, while the
other is the invariant random matrix ensemble with a shallow,
log-square confinement potential. We demonstrate a close
correspondence between the spectral as well as eigenfuncton
statistics of these random matrix ensembles and those of the random
tight-binding Hamiltonian in the point of the Anderson localization
transition in three dimensions. Finally we present a simple field
theory in 1+1 dimensions which reproduces level statistics of both
of these random matrix models and the classical Wigner-Dyson
spectral statistics in the framework of the unified formalism of
Luttinger liquid. We show that the (equal-time) density correlations
in both random matrix models correspond to the finite-temperature
density correlations of the Luttinger liquid. We also  present a
mechanism of the finite-temperature generation with breaking the
translational invariance in space by a metric with the event
horizon, similar to the problem of Hawking radiation in the black
holes.

\sect{Introduction}\label{intro} It was known since the pioneer's
work by F.Wegner \cite{Wegner1980} that the eigenfunctions
$\psi_{i}({\bf r})$ of the random Schroedinger operator
$\hat{H}\psi_{i}({\bf r})=E_{i}\,\psi_{i}({\bf r})$ at the mobility
edge $E_{i}=E_{m}$ corresponding to the critical point of the
Anderson localization transition, possess the property of
multifractality. In particular, at $E=E_{m}$ the moments of the
inverse participation ratio:
\begin{equation}
\label{P-q} P_{n}(E)=\sum_{{\bf r}}\sum_{i}\langle|\psi_{i}({\bf
r})|^{2n}\,\delta(E_{i}-E)\rangle\propto L^{-d_{n}(n-1)},
\end{equation}
scale as a certain power-law with the total size $L$ of the system.
This power is a critical exponent which depends only on the basic
symmetry (see Ref.\cite{Mehta} and Chapter 3 of this book) of the
Hamiltonian $\hat{H}$ and on the dimensionality of space $d$. The
true extended states of the Schroedinger operator are characterized
by all $d_{n}=d$. This allows to interpret $d_{n}<d$ as a certain
{\it fractal dimension} which depends on the order $n$ of the
moment. As a matter of fact the statistics of critical
eigenfunctions are described by a {\it set} of fractal dimensions
$d_{n}$ which justifies the notion of {\it multi}-fractality.

Another aspect of criticality is   the scaling with  the energy
difference $\omega=|E_{i}-E_{j}|$ between {\it two}  eigenvalues. It
is similar to the {\it dynamical scaling} and is relevant for the
correlation functions of {\it different} eigenfunctions
$\psi_{i}({\bf r})$ and $\psi_{j}({\bf r})$. The most important of
them is the local density of states correlation function:
\begin{equation}
\label{LDoS} C(\omega,R)=\sum_{{\bf r}}\sum_{i\neq
j}\langle|\psi_{i}({\bf r})|^{2}\,|\psi_{j}({\bf
r+R})|^{2}\,\delta(E-E_{i})\delta(E+\omega-E_{j})\rangle.
\end{equation}
This correlation function is relevant for the matrix elements of the
two-body interaction. The  dynamical scaling connects the power law
behavior of $C(\omega,0)\propto \omega^{-\mu}$ with that of
$C(0,R)\propto R^{-(d-d_{2})}$ by a conjecture \cite{Chalker} on the
dynamical exponent:
\begin{equation}\label{Chalker-anzats}
R^{d}\rightarrow \omega,\;\;\;\;\;\mu=1-\frac{d_{2}}{d}.
\end{equation}
Although there is an extensive numerical evidence in favor of
conjecture Eq.(\ref{Chalker-anzats})  its rigorous proof has been
lacking so far.

Last but not least, there is a growing interest to the spectral
(level) statistics in quantum systems whose classical counterparts
are in between of chaos and integrability \cite{Bogomolny}, the
simplest of them being the {\it two-level spectral correlation
function} (TLSCF) $R(\omega)\propto \sum_{{\bf R}}C(\omega,R)$. Some
of them apparently share the characteristic features of spectral
statistics at the Anderson localization transition, e.g. the finite
level compressibility $0<\chi<1$ \cite{Kot,CKL}, and the Poisson
tail of the level spacing distribution function $P(s)\propto
e^{-s/2\chi}$ combined with the Wigner-Dyson level repulsion
$P(s)\propto s^{\beta}$ at small level separation $s\ll 1$
\cite{Shklovskii}. There is a conjecture that these features are
also related with the multifractality of critical eigenfunctions,
however its exact formulation has not been developed beyond the
limit of weak multifractality $\chi\ll 1$, where one can show that
$\chi=\frac{1}{2}(1-d_{2}/d)$ \cite{CKL}.

The main reason for  scarcity of rigorous  knowledge about
multifractality of critical eigenfunctions is a lack of exactly
solvable models of sufficient generality. The most popular
three-dimensional (3D) Anderson model of localization
\cite{Anderson} is quite efficient for numerical simulations but so
far evaded any rigorous analytical treatment in the critical region.
More perspective seemed to be the Chalker's network model for the
Quantum Hall transition \cite{Chalker-network} and its
generalizations. However, numerous proposals for the critical field
theory were not successful so far \cite{parabola}. In this situation
the random matrix theory may prove to be the simplest, universal and
representative tool to obtain a rigorous knowledge about the
critical eigenfunction and spectral statistics \cite{KM}.

\sect{Non-invariant Gaussian random matrix theory with multifractal
eigenvectors.} Guided by the idea that multifractality of
eigenstates is the hallmark of criticality, we introduce the
Gaussian random matrix ensemble \cite{MF,KM} which eigenvectors obey
Eq.(\ref{P-q}) with $L$ being replaced by the matrix size $N$. This
random matrix theory and its modifications describes very well not
only the critical eigenfunction statistics at the Anderson
localization transition in three-dimensional (3D) Anderson model but
also the off-critical states close to the transition \cite{KC}. The
critical random matrix ensemble (CRMT) suggested in \cite{MF,KM} is
{\it manifest non-invariant}, and is defined as follows:
\begin{equation}
\label{cRMT} \langle H_{nm} \rangle=0,\;\;\;\;\;\langle
|H_{nm}|^{2}\rangle = \left\{\begin{matrix}\beta^{-1},& n=m\cr
\frac{1}{2}\,\left[ 1+\frac{(n-m)^{2}}{b^{2}}\right]^{-1}, & n\neq m
\end{matrix} \right.
\end{equation}
where $H_{nm}$ is the Hermitean $N\times N$ random matrix with
entries $H_{nm}$ ($n>m$) being  independent Gaussian random
variables; $\beta=1,2,4$ for the Dyson orthogonal, unitary and
symplectic symmetry classes, and $b>0$ is the control parameter.
This CRMT can be considered as a {\it particular deformation} of the
Wigner-Dyson RMT which corresponds to $b=\infty$.

As is clear from the definition Eq.(\ref{cRMT}) the  variance
$\langle |H_{nm}|^{2}\rangle$ is non-invariant under unitary
transformation $\hat{H}\rightarrow U\,\hat{H}\,U^{\dagger}$. The
existence of the preferential basis is natural, as this RMT mimics
the properties of the Anderson model of {\it localization} which
happens in the {\it co-ordinate space}, and not necessarily e.g. in
the momentum space.

The critical nature of the CRMT is encoded in the {\it power-law}
decay of the variance matrix, the typical off-diagonal entry being
proportional to $|n-m|^{-1}$ in the absolute value. This is exactly
the decay with the power equal to the dimensionality of space ($d=1$
in the case of matrices). In contrast to the Wigner-Dyson RMT which
probability distribution is parameter-free, the CRMT is a {\it
one-parameter family}. The parameter $b$ controls the spectrum of
fractal dimensions $d_{n}=d_{n}(b)$. One can show \cite{MF,ME} that
both at $b\gg 1$ and $b\ll 1$ the scaling relation Eq.(\ref{P-q})
holds true, and the basic fractal dimension is equal to:
\begin{equation}
\label{fr-b} d_{2}=\left\{\begin{matrix}1- c_{\beta}\,B^{-1},& B\gg
1 \cr c_{\beta}\,B, & B\ll 1
\end{matrix}\right.
\end{equation}
where
$c_{\beta}=\frac{\pi^{\frac{\beta}{2}-1}}{\beta}\,2^{\frac{1}{4}}$,
and $B=b\,\pi^{\frac{\beta}{2}}\,2^{\frac{1}{4}}$.

Eq.(\ref{fr-b}) can be cast in a form of the {\it duality
relationship}:
\begin{equation}
\label{dual} d_{2}(B)+d_{2}(B^{-1})=1.
\end{equation}
\begin{figure}[h]
\unitlength1cm
\begin{center}
\begin{picture}(12.2,8.5)
\epsfig{figure=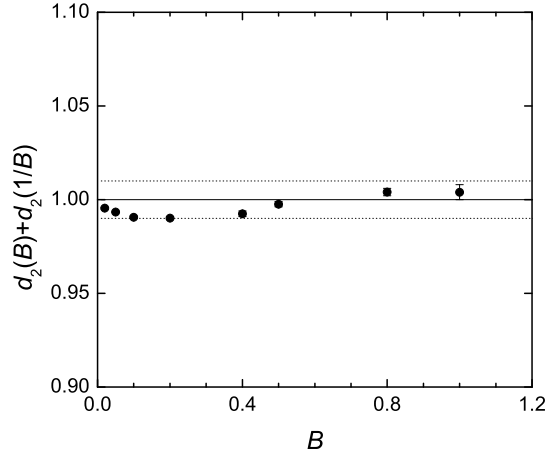, width=20pc}
\end{picture}
\end{center}
\caption{Numerical verification of the duality relation
Eq.(\ref{dual}), done by the box counting method \cite{KYOC} for the
unitary CRMT ensemble Eq.(\ref{cRMT}). The deviations from $1$ in
Eq.(\ref{dual}) do not exceed $1\%$ in the entire region of
$B\in[0,1]$.}
\end{figure}
The duality relation has been checked numerically \cite{KYOC} for
the unitary CRMT Eq.(\ref{cRMT}). The results are presented in
Fig.1.1. In particular, it was found that $2d_{2}(B=1)=1.003\pm
0.004$. The fact that the deviation of the sum $d_{2}(B)+d_{2}(1/B)$
from $1$ does not exceed $1\%$ in the entire region of $B\in[0,1]$
looks extremely interesting. However, it is not known yet whether an
exact function $d_{2}(B)$ obeys this remarkable duality relationship
which suggests that $d_{2}=\frac{1}{2}$ at
$b=\frac{1}{\sqrt{\sqrt{2}\pi^{\beta}}}$.

There was vast numerical evidence that the CRMT Eq.(\ref{cRMT})
reproduces the main qualitative features of the critical
eigenfunction and spectral statistics, including the power-law
behavior of the DoS correlation function $C(\omega,0)$ [see
Ref.\cite{KC} and references therein], distributions of moments of
inverse participation ratio $P_{q}$ \cite{ME}, role of rare
realizations and multifractality spectrum close to termination point
\cite{Mirlin-rep}, and the hybrid Wigner-Dyson \& Poisson level
spacing distribution \cite{Shklovskii}.

The very fact that by a choice of control parameter $B$ one can fit
quite accurately
the critical
statistics of both eigenvalues and eigenfunctions of the Anderson
localization model, is extremely encouraging.

\sect{Invariant RMT with log-square confinement} Quite remarkably,
there is an {\it invariant} (but non-Gaussian) RMT whose {\it
two-level spectral correlation function} $R_{N}(s,s')$ is closely
related with that of the CRMT discussed in the previous section.
Namely, in the unfolded energy variables $s$ in which the mean
spectral density ("global Density of States") is equal to unity, and
in the large $B$ limit one finds:
\begin{equation}
\label{tr-inv-DoS}  R_{\infty}(s-s')\left|_{{\rm inv.}} \right.=
R_{\infty}(s-s')\left|_{{\rm non-inv.}}  \right. ,
\end{equation}
where $R_{\infty}(s-s')=\lim_{s\rightarrow\infty}(\lim_{N\rightarrow
\infty} R_{N}(s,s'))$.

For the unitary symmetry class one can show \cite{KM} that:
\begin{equation}
\label{u-TLCF} R_{\infty}(s-s')\left| _{{\rm
non-inv}}\right.=1-\pi^{2}\kappa^{2}\,\frac{
\sin^{2}(\pi(s-s'))}{\sinh^{2}[\pi^{2}\kappa(s-s')]},\;\;\;\;\;\kappa=\frac{\beta}{2}\,\chi(b),
\end{equation}
where $\kappa$ is related with the level compressibility $\chi(b)$
(see Eq.(\ref{SR-vio}),(\ref{exp-P}) below). This suggests
\cite{Mehta} the form of the kernel $K(s-s')$ of the invariant RMT
which is the only input one needs to compute all many-point spectral
correlation functions:
\begin{equation}
\label{kern} K(s-s')=\pi\kappa\,\frac{
\sin(\pi(s-s'))}{\sinh[\pi^{2}\kappa(s-s')]}.
\end{equation}
Thus a remarkable correspondence between an invariant and a
non-invariant RMT can be conjectured \cite{KM} which allows to use
the full power of the unitary invariance for calculation of spectral
statistics.

Now it is time to specify the invariant RMT whose counterpart in
Eq.(\ref{tr-inv-DoS}) is the critical RMT with multifractal
eigenstates. The probability distribution for the random matrix
Hamiltonian $\hat{H}$ is \cite{Muttalib, KM}:
\begin{equation}
\label{mutta} P(\hat{H})\propto {\rm exp}\left[-\beta\,{\rm
tr}\,V(\hat{H})\right],\;\;\;\;V(x)=\sum_{n=0}^{\infty}\ln\left[
1+2q^{n+1}\,(1+2x^{2})+q^{2(n+1)}\right],
\end{equation}
where $1<q<0$ is a control parameter.

It is extremely important that the "confinement potential" $V(x)$ at
large $|x|$ behaves like:
\begin{equation}
\label{log-2} V(x)\rightarrow
A\,\ln^{2}|x|,\;\;\;\;\;A=\frac{2}{\ln(q^{-1})}.
\end{equation}
thus being an example of a "shallow" confinement. The correspondence
Eq.(\ref{tr-inv-DoS}) holds for $\ln q^{-1}<<1 $, while for $\ln
q^{-1}>\pi^{2}$ an interesting phenomenon of "crystallization" of
eigenvalues happens with the TLCF taking a triangular form
\cite{Muttalib,KM,Bogomolny-Pato}.

The form of the confinement potential in Eq.(\ref{mutta}) is quite
specific, even within the class of shallow potentials with a
double-logarithmic asymptotic behavior Eq.(\ref{log-2}). It has been
chosen in Ref.\cite{Muttalib} in order to enable an exact solution
in terms of the $q$-deformed Hermite polynomials. However, there
exists another, more simple argument why this particular form leads
to an exact solution \cite{Sedrakyan}. The point is that the measure
Eq.(\ref{mutta}) can be considered as a {\it generalized Cauchy
distribution}:
\begin{equation}
\label{Cauchy}
V(E_{i})=\prod_{n=1}^{\infty}\frac{1}{4q^{n+1}}\prod_{i=1}^{N}
\frac{1}{E_{i}^{2}+\Gamma_{n}^{2}},\;\;\;\;\;\Gamma_{n}=\frac{1+q^{n+1}}{2q\frac{n+1}{2}}.
\end{equation}
\sect{Self-unfolding and not self-unfolding invariant RMT.} As we
will see below, the cause of all the peculiarities of the RMT with
double-logarithmic confinement is the fact that the mean density of
states
$\rho_{\infty}(E)=\lim_{N\rightarrow\infty}\sum_{i=1}^{N}\langle
\delta(E-E_{i}) \rangle$ approaches some stable {\it non-trivial}
form in the bulk of the spectrum as the size of matrix $N$ tends to
infinity. This is in contrast to the Wigner-Dyson classical RMT
where $\lim_{N\rightarrow\infty}\rho_{N}(E)/\sqrt{2N}=1$ is
independent of $E$. This is the reason why the Wigner-Dyson RMT can
be referred to as {\it self-unfolding}, while the RMT with the
double-logarithmic confinement is {\it not self-unfolding}.

In order to find a criterion for an invariant RMT to be
self-unfolding, let us apply the Wigner-Dyson plasma analogy
\cite{Mehta}. According to this approximation the mean density of
states $\rho_{N}(E)$ obeys the integral equation of the equilibrium
classical plasma with logarithmic interaction subject to the
confining force $-dV/dE$:
\begin{equation}
\label{plasma}
v.p.\int_{-\infty}^{+\infty}\rho_{N}(E')\,\frac{dE'}{E-E'}=\frac{dV}{dE}\equiv
f(E),
\end{equation}
where $v.p.$ denotes the principle value of the integral. The
solution to this equation in the case of symmetric confining
potential is:
\begin{equation}
\label{sol}
\rho_{N}(E)=\frac{1}{\pi^{2}}\,\sqrt{D_{N}^{2}-E^{2}}\,\,v.p.\int_{-D_{N}}^{D_{N}}\frac{f(E')}{\sqrt{D_{N}^{2}-E'^{2}}}
\,\frac{dE'}{E'-E},
\end{equation}
where the band-edge $D_{N}$ should be determined from the
normalization condition $\int_{-D_{N}}^{D_{N}} \rho_{N}(E')\,dE'=N$.

One can see that if $|f(E')|$ increases slower than $|E'|$ as
$|E'|\rightarrow\infty$, the integral converges to a non-trivial
function as $N$ and $D_{N}\rightarrow\infty$:
\begin{equation}
\label{sol-non-un}
\rho_{\infty}(E)=\frac{1}{\pi^{2}}\,\,v.p.\int_{-\infty}^{+\infty}dE'\,\frac{f(E')}{E'-E}.
\end{equation}
Otherwise, the large values of $|E'|\sim D_{N}$ dominate the
integral in Eq.(\ref{sol}), so that at a fixed $E$ and
$N\rightarrow\infty$ (when $D_{N}\gg |E|$) the dependence of
$\rho_{N}(E)$ on $E$ disappears. We conclude therefore \cite{KrCan,
Can} that the criterion of a {\it non-self-unfolding} RMT is:
\begin{equation}
\label{cr-self} \lim_{|x|\rightarrow\infty} \frac{V(x)}{|x|}=0.
\end{equation}
Let us consider the {\it shallow} confining potential with the
power-law large-$x$ asymptotic behavior
\begin{equation}
\label{V-alpha} V(x)=A|x|^{\alpha},\;\;\;\; (\alpha<1).
\end{equation}
 Then at large $|E|\gg 1$ we have
\cite{KrCan,Can}:
\begin{equation}
\label{pow} \rho_{\infty}(E)=\frac{A\alpha}{\pi}\,\tan\left(
\frac{\pi\alpha}{2}\right)\,\frac{1}{|E|^{1-\alpha}}.
\end{equation}
In particular for the log-square confining potential \footnote{For
$V(x)\propto \ln^{d}|x|$ with $d\geq 2$ the leading term at large
$E$ is $\rho_{\infty}(E)\propto \frac{\ln^{d-2}|E|}{|E|}$. }
\begin{equation}
\label{ln-2-pow} V(x)=A\ln^{2}|x|=A\lim_{\alpha\rightarrow 0}\left(
\frac{|x|^{\alpha}-1}{\alpha}\right)^{2} =A\,\lim_{\alpha\rightarrow
0}\frac{|x|^{2\alpha}-2|x|^{\alpha}+1}{\alpha^{2}}.
\end{equation}
one finds using the linear dependence of Eq.(\ref{sol-non-un})
$\rho_{\infty}(E)$ on $f(E)$ and Eq.(\ref{V-alpha}):
\begin{equation}
\label{d-l-DoS} \rho_{\infty}(E)=\frac{A}{|E|}.
\end{equation}
There is a qualitative and far-reaching difference between the
shallow power-law confinement potential with $0<\alpha<1$, and the
log-square confinement. Although both lead to a non-self-unfolding
RMT, the case of finite $\alpha$ can be called {\it weakly non-self
unfolding}, because for large enough $E$ the variation of the mean
density at a scale of the mean level spacing
$\Delta=\rho_{\infty}^{-1}$ is much smaller than the density itself:
\begin{equation}
\label{weak n-a}
\frac{\rho_{\infty}(E+\Delta)-\rho_{\infty}(E)}{\rho_{\infty}(E)}=\frac{\rho_{\infty}'}{\rho_{\infty}^{2}}\propto
|E|^{-\alpha}\rightarrow 0.
\end{equation}
In contrast to that, the log-square confinement is {\it strongly
non-self unfolding}, since the ratio in Eq.(\ref{weak n-a}) is
always finite \footnote{We put the confinement potentials
$\ln^{d}|E|$ with $d>2$ to the class of strongly non-self unfolding
potentials even though the ratio Eq.(\ref{weak n-a}) logarithmically
vanishes at large $E$.}.

We conclude that depending on the steepness of the confinement
potential at large $E$ there are three qualitatively different cases
\cite{KrCan,Can}:
\begin{itemize}
\item
{\it self-unfolding RMT} for $\lim_{x\rightarrow \infty}V(x)/|x|>0$
\item
{\it weakly non-self-unfolding RMT} for $\lim_{x\rightarrow
\infty}V(x)/|x|=0$ but $\exists\alpha>0$ such that
$\lim_{x\rightarrow \infty}V(x)/|x|^{\alpha}>0$
\item
{\it strongly non-self unfolding RMT} if $\forall \alpha>0$ holds
$\lim_{x\rightarrow \infty}V(x)/|x|^{\alpha}=0$
\end{itemize}
The first case is characterized by the Wigner-Dyson universality of
the spectral correlation functions. In the second case this
universality holds only approximately for sufficiently large
distance from the origin, while if one or two energies are close to
the origin, the Wigner-Dyson universality is no longer valid, even
after unfolding. In the third case, the Wigner-Dyson universality is
not valid also in the bulk of the spectrum.

\sect{Unfolding and the spectral correlations} Let us consider the
power-law confinement Eq.(\ref{V-alpha}) and find the corresponding
{\it unfolding co-ordinates} $s(E)$ in which the spectral density is
1:
\begin{equation}
\label{unfold} s(E)={\rm
sgn}(E)\,\int_{0}^{E}\rho_{\infty}(E')\,dE'=\frac{A}{\pi}\,\tan\left(
\frac{\pi\alpha}{2}\right)\;{\rm sgn}(E)\,|E|^{\alpha}.
\end{equation}
Note that for large enough $E$ the unfolding co-ordinates are given
by Eq.(\ref{unfold}) even if $V(E)$ is not a pure power-law at small
$E$ but is rather deformed to have a regular behavior at the origin.

The corresponding unfolding co-ordinates for the log-square
potential Eq.(\ref{ln-2-pow}) are:
\begin{equation}
\label{unfold-ln} s(E)=A\;{\rm sgn}(E)\;\ln(c|E|),\;\;\;\;cE(s)={\rm
sgn}(s)\;e^{\frac{|s|}{A}}.
\end{equation}
where $c$ is a constant which depends on the regularization of the
log-square potential   close to the origin.

As we will see, the exponential change of co-ordinates
Eq.(\ref{unfold-ln}) leads to dramatic consequences for two-level
spectral correlations, and  has even a far-reaching analogy in the
physics of black holes \cite{KrCan, Fabio}.

In order to show how a non-trivial unfolding changes the form of the
spectral correlations consider the spectral kernel $K_{N}(E,E')$
given in terms of the orthogonal polynomials by the
Christoffel-Darboux formula \cite{Mehta}:
\begin{equation}
\label{Kr-Dar}
K_{N}(E,E')=\left|\frac{\varphi_{N-1}(E)\varphi_{N}(E')-\varphi_{N-1}(E')\varphi_{N}(E)}{E-E'}\right|,
\end{equation}
where $\varphi_{N}(E)$ is the (properly normalized) "wave-function"
which is related with the orthogonal polynomials $p_{n}$:
\begin{equation}
\label{wf}
\varphi_{N}(E)=p_{n}(E)\,e^{-\beta\,V(E)/2},\;\;\;\;\;\int_{-\infty}^{+\infty}p_{n}(E')\,p_{m}(E')\,e^{-\beta\,
V(E')}\,dE'=\delta_{nm}.
\end{equation}
Any spectral correlation function can be expressed \cite{Mehta} in
terms of the kernel Eq.(\ref{Kr-Dar}). In particular, the $n$-point
Density of States correlation function at $\beta=2$ takes the form
\cite{Mehta}:
\begin{equation}
\label{n-point} R_{\infty}(E_{1},...E_{n})={\rm
det}\,\left[K_{\infty}(E_{i},E_{j})\right]\,\prod_{i=1}^{n}K^{-1}(E_{i},E_{i}).
\end{equation}
Below we derive a {\it semi-classical} form of the kernel which is
valid in the $N\rightarrow \infty$ limit provided that the
coefficient $A$ in Eqs.(\ref{V-alpha}),(\ref{ln-2-pow}) is large. In
particular for the RMT with log-square confinement given by
Eq.(\ref{mutta}) the condition of applicability of the analysis done
below is $\ln q^{-1}<2\pi$ \cite{Muttalib}.

In  this semiclassical limit the "wave functions" take the form (we
assume that the confinement potential is an even function of $E$):
\begin{equation}
\label{sem} \varphi_{N-1}(E)=\sin(\pi s(E)),\;\;\;\;\;
\varphi_{N}=\cos(\pi s(E)),
\end{equation}
if $N$ is even and $\cos\rightarrow \sin$ if $N$ is odd. Then one
finds in the unfolding co-ordinates:
\begin{equation}
\label{K-inf} K_{\infty}(E,E')=\frac{\sin(\pi(s-s'))}{E(s)-E(s')}.
\end{equation}
Equation Eq.(\ref{K-inf}) demonstrates that as long as the
semiclassical approach applies, the non-trivial unfolding $E=E(s)$
is the only source of deformation of the kernel and its deviation
from the universal Wigner-Dyson form.

The main conclusion one can draw from Eq.(\ref{K-inf}) is that the
translational invariance is lost in the $N\rightarrow\infty$ limit
for all non-self-unfolding invariant RMT. It is not sufficient to
have the mean density $\rho_{\infty}(s)$ equal to unity for all
values of $s$ in order to have all the correlation functions of the
universal Wigner-Dyson form.

\sect{Ghost correlation dip in RMT and Hawking radiation. } The
break-down of translational invariance takes especially dramatic
form in case of the log-square confinement. We will show below  that
in this case a {\it ghost correlation dip} appears in the two-level
spectral correlation function which position at $s\approx -s'$ is
{\it mirror reflected} relative to the position of the {\it
translational-invariant correlation dip} at $s\approx s'$.

Indeed let us consider the semiclassical kernel Eq.(\ref{K-inf})
with the exponential unfolding Eq.(\ref{unfold-ln})  for $s\gg 1$
and $|s-s'|<<|s|$. Then after a simple algebra we obtain for
$s\,s'>0$ the Two-level Spectral Correlation Function given by
Eq.(\ref{u-TLCF}) $(\beta=2)$:
\begin{equation}
\label{kappa-A} R_{\infty}(s\,s'>0)=1-\pi^{2}\kappa^{2}\,\frac{
\sin^{2}(\pi(s-s'))}{\sinh^{2}[\pi^{2}\kappa(s-s')]},\;\;\;\;\;
\kappa=\frac{1}{2\pi^{2}\,A}=\frac{\ln q^{-1}}{4\pi^{2}}\ll 1.
\end{equation}
\begin{figure}[h]
\unitlength1cm
\begin{center}
\begin{picture}(10.2,10.5)
\epsfig{figure=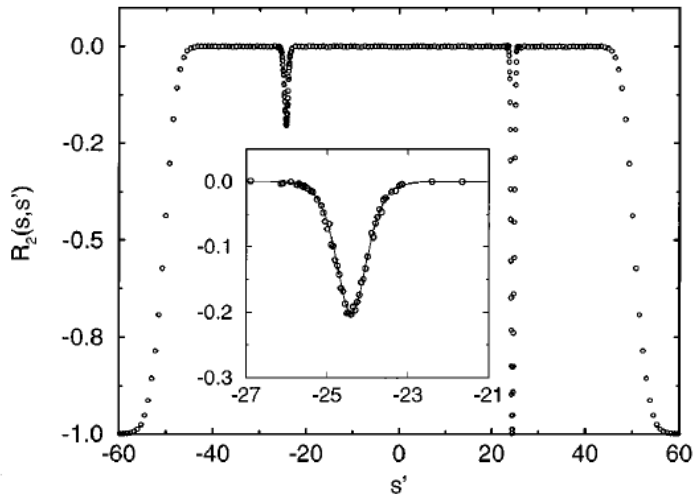, width=30pc}
\end{picture}
\end{center}
\caption{The irreducible part $R_{\infty}(s,s')-1$ of the TLSCF
obtained \cite{Can} by the classical Monte-Carlo simulations at a
temperature $T=\frac{1}{2}$ on the one-dimensional plasma with
logarithmic interaction   in the log-squared confinement potential
Eq.(\ref{ln-2-pow}) with $A=0.5$. Two dips of the anti-correlations
are well seen with the smaller {\it ghost dip} shown in detail in
the insert. The edges of the correlation function correspond to the
finite number of particles (energy levels) $N=101$.}
\end{figure}
This correlation functions exhibits a dip of anti-correlation (level
repulsion) at small $s-s'$  and approaches exponentially fast the
uncorrelated asymptotic  $R_{\infty}(s,s')=1$ for $|s-s'|>
1/\pi^{2}\kappa$.  An amazing fact is that the anti-correlation
revives again when $s+s'$ becomes small. Considering $s\,s'<0$ and
plugging the exponential unfolding Eq.(\ref{unfold-ln}) into
Eq.(\ref{K-inf}) we obtain for the unitary case $\beta=2$:
\begin{equation}
\label{ghost}
R_{\infty}(s\,s'<0)=1-\pi^{2}\kappa^{2}\,\frac{\sin^{2}(\pi(s-s'))}{\cosh^{2}(\pi^{2}\kappa\,(s+s'))}.
\end{equation}
This is the {\it ghost anti-correlation peak} discovered in
Ref.\cite{KrCan} but also present in the exact solution
\cite{Muttalib}.

The existence of such a ghost correlation dip is requested by the
normalization sum rule \cite{KrCan} which for the invariant RMT
survives taking the limit $N\rightarrow\infty$:
\begin{equation}
\label{nsr} \int_{-\infty}^{+\infty}(1-R_{\infty}(s,s'))\,ds'=1.
\end{equation}
Substituting in Eq.(\ref{nsr}) the sum of the translational
invariant and the ghost peak  found from
Eqs.(\ref{kappa-A}),(\ref{ghost}) and doing the integral one
obtains:
\begin{equation}
\label{acc}
\coth(\kappa^{-1})-\frac{1}{\sinh(\kappa^{-1})}\,\cos(4\pi\,
s)\approx 1.
\end{equation}
This expression  is equal to 1 with the exponential accuracy in
$e^{-\frac{1}{\kappa}}\ll 1$, which is exactly the accuracy of the
semiclassical approximation Eqs.(\ref{kappa-A}),(\ref{ghost}).

\begin{figure}[h]
\unitlength1cm
\begin{center}
\begin{picture}(12.2,8.5)
\epsfig{figure=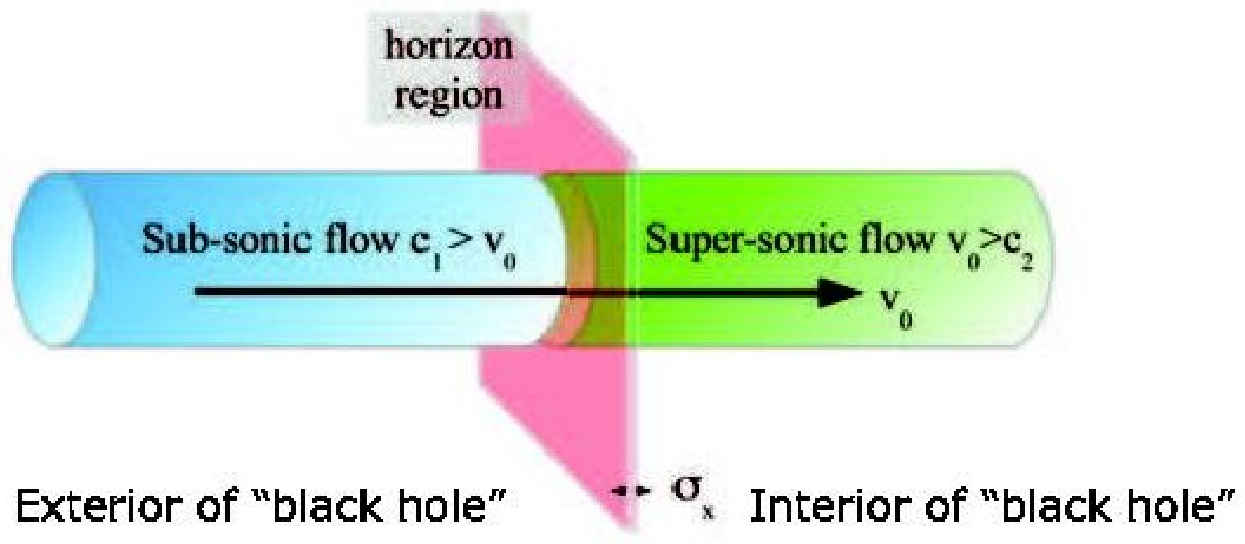, width=30pc}
\end{picture}
\end{center}
\caption{The sonic analogue of a black hole (courtesy of
I.Carusotto): the stream of cold atoms with the the velocity $v_{0}$
and the phonon velocity $c(x)$ changing near the origin from the
value $c_{1}>v_{0}$ to the value $c_{2}<v_{0}$. This region of the
super-sonic flow is analogous to the interior of the black hole
where light cannot escape from.}
\end{figure}

It appears that the situation with the ghost correlation dip has an
important physical realization \cite{Fabio}. Let us consider the
{\it sonic} analogue of the black hole \cite{Unruh, Caruzotto}
described in Fig.1.3. The stream of cold atoms moves with the
constant velocity $v_{0}$, while the interaction between atoms is
tuned so that the sound velocity is larger than the stream velocity
for $x<-\delta$ and smaller than the stream velocity for $x>\delta$.
The latter region where no phonon may escape from, is analogous to
the interior of a black hole. The point in space where $c(x)=v_{0}$
is analogous to the {\it horizon} in the black hole physics and the
phonon radiation emerging when the horizon is formed, is analogous
to the Hawking radiation \cite{Hawking}.

As it was shown in Ref.\cite{Caruzotto}, the Hawking radiation
possesses a peculiar correlation of photons (phonons): despite being
bosons, they statistically {\it repel} each other forming a dip in
the density correlation function not only at $x\approx x'$ but also
near the mirror point $x\approx -x'$, with the envelopes of the
normal and ghost dips proportional to $\sinh^{-2}$ and $\cosh^{-2}$,
respectively. As in the case of RMT with log-square confinement, the
ghost correlation dip appears in the absence of any symmetry that
would require the eigenvalues $E_{i}$ or the Hawking bosons to
appear in pairs simultaneously at  points $\pm E_{i}$ ($\pm x_{i}$).
More close inspection shows that the mechanism of its appearance in
quantum gravity and in the random matrix theory is very similar: it
is the exponential change of variables similar to
Eq.(\ref{unfold-ln}) which is aimed to make flat the mean DoS
(unfolding) in RMT or to make flat the metric of space-time in the
quantum gravity (exponential red-shift) \cite{Fabio}. Thus formation
of the strong non-self-unfolding RMT with the  log-square
confinement appears to be analogous to formation of a horizon in the
general relativity.

\sect{Invariant-noninvariant correspondence.} Note that the
translational-invariant part of spectral correlations described by
Eq.(\ref{kappa-A}) is {\it well separated} from the translational
non-invariant ghost dip if $|s|\gg |s-s'|$. As correlations
exponentially decrease at $|s-s'|>(\pi^{2}\kappa)^{-1}$, the scale
separation takes place when $|s|\gg (\pi^{2}\kappa)^{-1}$.
Essentially what happens because of the scale separation is that
(like in a black hole) the world is divided in two parts, and in
each of them local spectral correlations (which are approximately
translational-invariant ) are not affected by the presence of a
"parallel world". Moreover, the local correlations in the {\it
invariant} RMT with log-square confinement and $\kappa\ll 1$ were
conjectured (for the two-level correlations this conjecture has been
proven in \cite{KM}) to be the same as in the {\it non-invariant}
critical RMT \cite{KM,KTs}. This conjecture is related with the idea
of the {\it spontaneous breaking of unitary invariance} in RMT with
shallow confinement \cite{KrCan}. Although this idea has never been
proven or even convincingly demonstrated numerically, it seems to be
the only physically reasonable cause of the invariant-noninvariant
correspondence like the one given by Eq.(\ref{tr-inv-DoS}).

The invariant-noninvariant correspondence allows to use the
invariant RMT with log-square confinement for computing spectral
correlations of {\it non-invariant} RMT considered in sec.1.2. The
idea of such calculations is to use the kernel in the form
Eq.(\ref{kern}) to be plugged into the conventional machinery of
invariant RMT \cite{Mehta}. Here we present some results of such
calculations taken from Ref.\cite{Nishigaki}.
\begin{figure}[h]
\unitlength1cm
\begin{center}
\begin{picture}(50.2,5.5)  \epsfig{figure=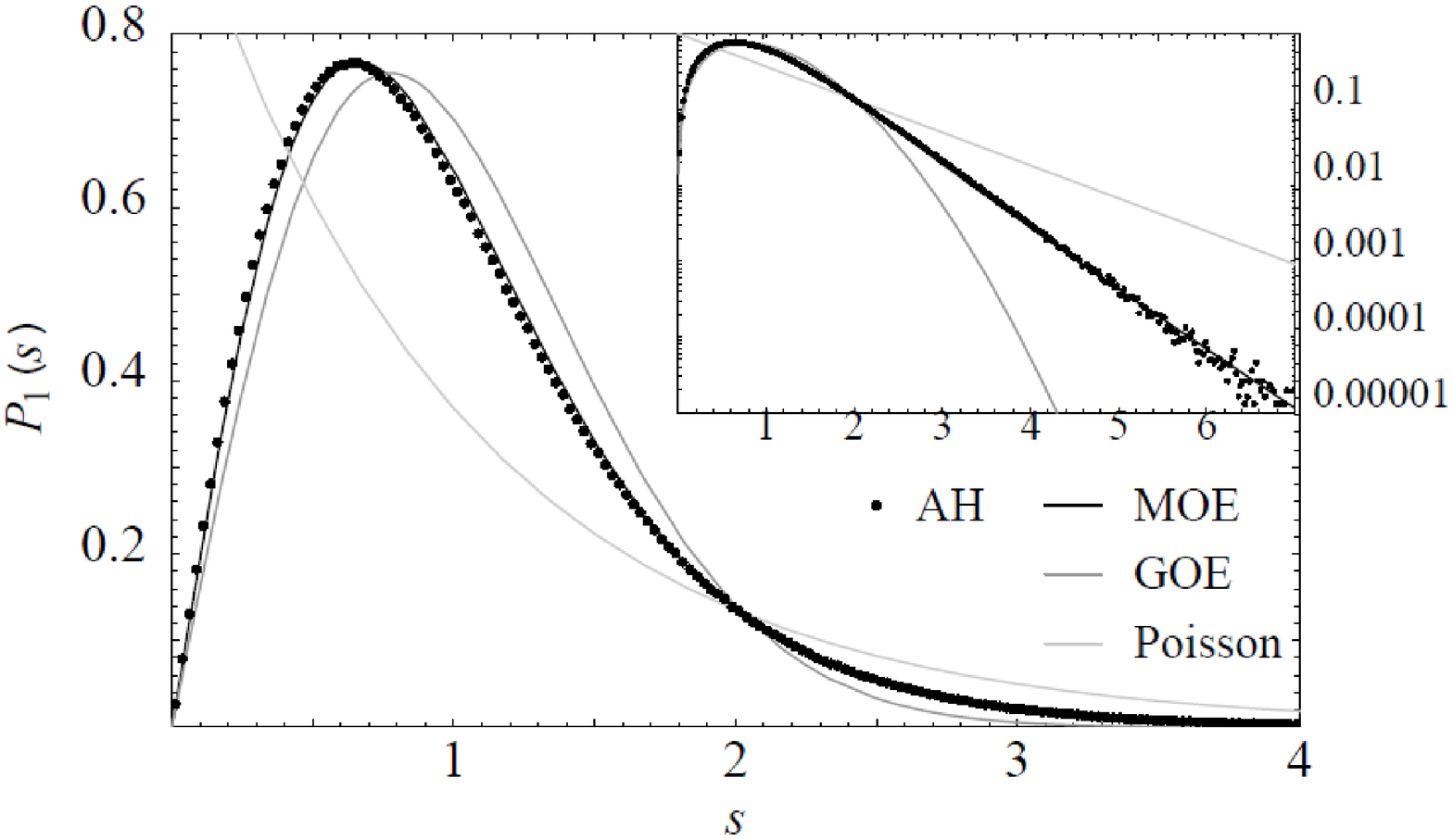, width=17pc}\epsfig{figure=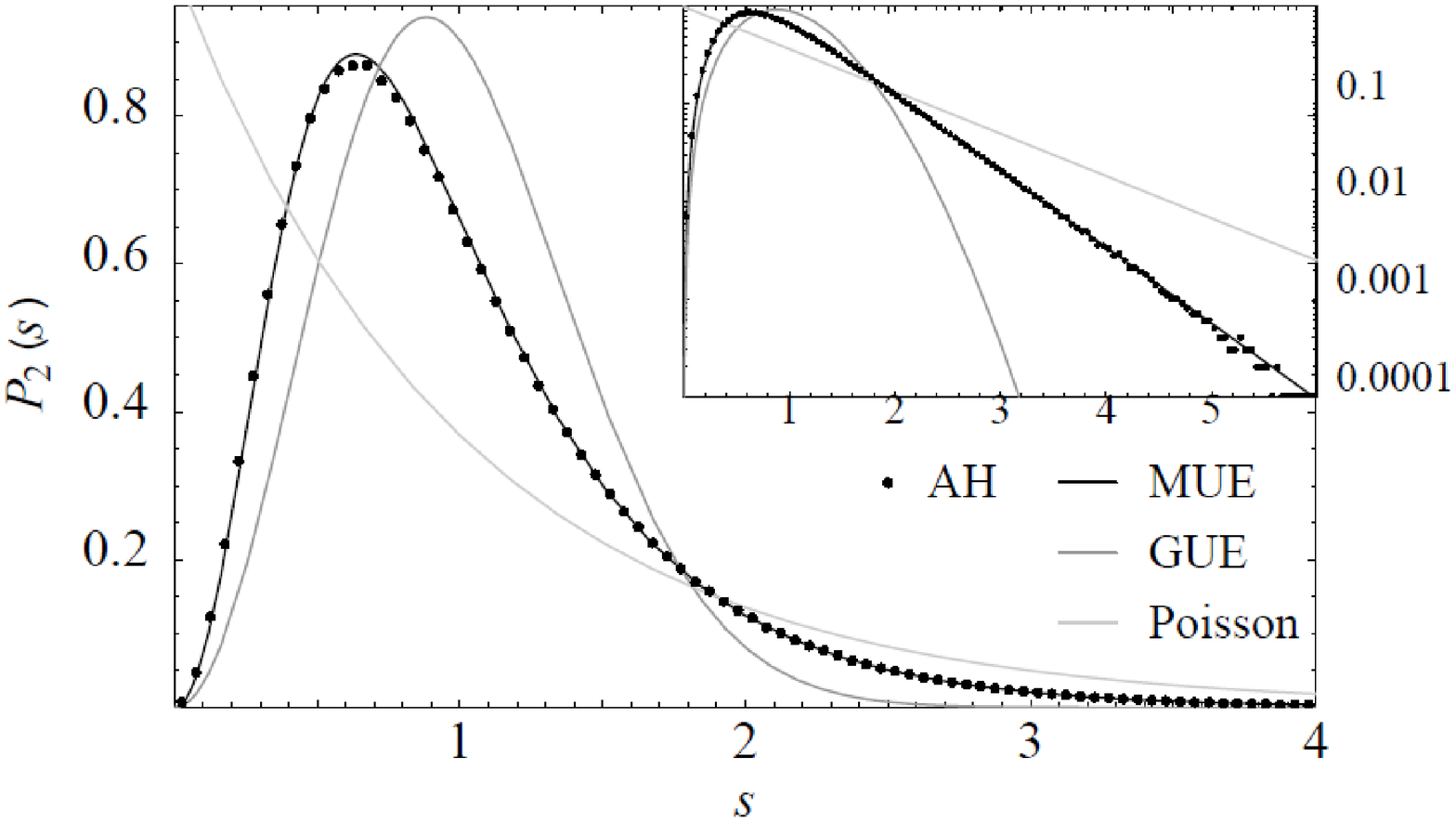,width=16pc}
\end{picture}
\end{center}
\caption{Level spacing distribution function (after Nishigaki
\cite{Nishigaki}) for the orthogonal and the unitary ensembles with
log-square confinement (black solid lines) and the corresponding
distributions for the 3D Anderson localization model with critical
disorder (data points). Grey solid lines are the Wigner-Dyson and
the Poisson distributions. The same values of $A=0.34$ (orthogonal
ensemble) and $A=0.14$ (unitary ensemble) allow to fit well both the
body and the far tail of the distribution shown in inserts.
}
\end{figure}
One can see from Fig.1.4 that the choice of one single parameter $A$
in the invariant RMT with log-square confinement allows to fit well
both the {\it body} and the {\it exponential tail} of the level
spacing distribution function obtained by numerical diagonalization
of the 3D Anderson localization model at criticality. The
corresponding parameters $b$ of the non-invariant CRMT
Eq.(\ref{cRMT}) should be chosen so that the parameter $\kappa$ in
the kernel Eq.(\ref{kern}) expressed through $A$ and $b$ is the
same. Comparing expressions for $\kappa$ in
Eqs.(\ref{u-TLCF}),(\ref{kappa-A}) we conclude that $b$ is a
solution to the equation:
\begin{equation}
\label{corresp} \frac{1}{\pi^{2}\,A}=\beta\,\chi(b).
\end{equation}
For the {\it unitary} ensemble the function $\chi(b)$ in the range
$\chi=(2\pi^{2}A)^{-1}\approx 0.36$ is well described \cite{KrNdaw}
by the large-$b$ asymptotic formula \cite{ME} $\chi(b)=(4\pi
b)^{-1}$. This gives an estimate $b=0.22$. Numerical diagonalization
of the CRMT shows that $b=0.22$ corresponds to $d_{2}\approx 0.46\pm
0.01$. To establish the correspondence with the Anderson
localization model of unitary symmetry at critical disorder we note
that the fractal dimension in the 1d CRMT corresponds to the {\it
reduced} dimension $d_{2}/3$ of the three-dimensional Anderson
model. Thus the fractal dimension of the Anderson model should be
compared with $0.46\times 3=1.38$. It appears to be in an excellent
agreement with the direct diagonalization of the Anderson model of
unitary symmetry which gives \cite{Cue-sing} $d_{2}=1.37$. This
example demonstrates that the values of parameters $A$ and $b$ found
from fitting the {\it spectral} statistics of RMT to that of the 3D
Anderson localization model, automatically give an excellent fitting
of the {\it eigenfunction} statistics.

 \sect{Normalization anomaly, Luttinger
liquid analogy and the Hawking temperature.} As has been already
mentioned, the local spectral correlations with $|s-s'|\ll |s|$ are
well described by the translational-invariant kernel
Eq.(\ref{kern}). An important difference between this kernel and
that of the conventional Wigner-Dyson theory is that it contains a
{\it second energy scale} $\kappa^{-1}$ in addition to the mean
level spacing $\Delta=1$. Moreover, the way this scale appears
through $s-s'\rightarrow \sinh(\pi^{2}\kappa\,(s-s'))$ suggests an
analogy with the
  system of one-dimensional
fermions at a {\it finite temperature}. Indeed, the density
correlations of non-interacting one-dimensional fermions (whose
ground state correlations are equivalent to the Wigner-Dyson
spectral statistics at $\beta=2$) is described by Eq.(\ref{u-TLCF}),
where
\begin{equation}
\label{T} T=\pi\,\kappa.
\end{equation}
Given the kernel Eq.(\ref{kern}) it is not difficult to obtain the
level density correlation functions for $\beta=1,4$ using the
standard formulae of the Wigner-Dyson RMT \cite{Mehta}. It appears
that both in the region of $|s-s'|\gg 1$ and in the region of
$|s-s'|\ll \kappa^{-1}$ they coincide  with the generalization of
Eq.(\ref{u-TLCF}) to $\beta=1,4$  obtained \cite{KTs}directly from
the {\it non-invariant} CRMT Eq.(\ref{cRMT}). For $\kappa\ll 1$
these regions overlap on a parametrical large interval, and one can
again demonstrate the invariant-noninvariant correspondence.

It is important to note that breaking the unitary invariance
explicitly by a finite $b$ in Eq.(\ref{cRMT}) leads to a {\it
normalization anomaly}. Namely, the normalization sum rule
\cite{KrCan} which exactly holds for the finite size of matrix $N$:
\begin{equation}
\label{SR-ex} 2\int_{s>0}^{+\infty}(1-R_{N}(s'-s))\,ds'=1,
\end{equation}
gets violated if the limit $N\rightarrow\infty$ is taken {\it prior}
to doing the integral. For instance at $\beta=2$  we have with
exponential accuracy $e^{-1/\kappa}\ll 1$:
\begin{equation}
\label{SR-vio}
\eta=2\int_{s>0}^{\infty}(1-R_{\infty}(s'-s))\,ds'=\coth(\kappa^{-1})-\kappa\approx
1-\kappa.
\end{equation}
For $\beta=1$ the corresponding expression is:
\begin{equation}
\label{SR-vio-1}
\eta=2\cot(\kappa^{-1})-\tanh(1/2\kappa)-2\kappa\approx 1-2\kappa.
\end{equation}

This leads to the finite {\it level compressibility} \cite{CKL}:
\begin{equation}
\label{compr} \chi\equiv\frac{d}{d \bar{n}}\,\langle
(n-\bar{n})^{2}\rangle=1-\eta,\;\;\;\;\;\;N\gg\bar{n}=\langle n
\rangle \gg 1,
\end{equation}
and the exponential (instead of the Gaussian in the Wigner-Dyson
RMT) tail of the level spacing distribution function:
\begin{equation}
\label{exp-P} \ln P(s)\approx -\frac{s}{2\chi}.
\end{equation}
Both properties Eq.(\ref{compr}) and Eq.(\ref{exp-P}) are the
signatures of criticality \cite{Shklovskii}. Due to
invariant-noninvariant correspondence the normalization anomaly
Eq.(\ref{SR-vio}) holds also for the invariant RMT with log-square
confinement thus again rasing a question on the {\it spontaneous
breaking of unitary invariance}.

In the absence of the formal proof of this conjecture, we present
here a simple theory which may {\it unify} both of the ensembles.
The idea of such a theory stems from the fact \cite{ALee} that the
classical Wigner-Dyson RMT \cite{Mehta} is equivalent to the {\it
ground state} of the one-dimensional Calogero-Sutherland model
\cite{Cal-Suth} of {\it fermions} with inverse square interaction of
the strength $\frac{\beta}{2}\,\left( \frac{\beta}{2}-1\right)$ in a
harmonic confinement potential. The large-scale properties of such a
model are described by the Luttinger liquid phenomenology
\cite{Haldane}. Its simplest finite-temperature formulation
\cite{Nersesyan} is in terms of the free-{\it bosonic} field in a
$1+1$ space-time
$\Phi(s,\tau)=\frac{1}{2}\,[\Phi_{R}(s,\tau)+\Phi_{L}(s,\tau)]$ with
the quadratic action:
\begin{equation}
\label{boson-act} S_{T}[\Phi]=\frac{\beta}{4\pi\,
K}\int_{0}^{1/T}d\tau\int_{\infty}^{+\infty}
ds\,[(\partial_{s}\Phi)^{2}+(\partial_{\tau}\Phi)^{2}],\;\;\;\;\;\Phi(s,\tau)=\Phi(s,\tau+1/T).
\end{equation}
The density correlation functions are given by the {\it functional
averages} of the density operator:
\begin{equation}
\label{DO}
\rho(s,\tau)=\rho_{0}+\frac{1}{\pi}\,\partial_{s}\Phi(s,\tau)+A_{1}\,\cos[2\pi\,s+2\Phi(s,\tau)]
+A_{2}\,\cos[4\pi\,s+4\Phi(s,\tau)]+...,
\end{equation}
where $A_{k}$ are structural constants which are determined by
details of interaction at small distances and take some fixed values
for the Calogero-Sutherland model corresponding to the symmetry
class $\beta$. Using Eq.(\ref{DO}) one may express the density
correlation functions in terms of the fundamental Green's function:
\begin{equation}
\label{funGF}
\langle\Phi(s,\tau)\,\Phi(s',\tau)\rangle_{S}-\frac{1}{2}\langle\Phi(s,\tau)\,\Phi(s,\tau)\rangle_{S}-\frac{1}{2}
\langle\Phi(s',\tau))\,\Phi(s',\tau)\rangle_{S}=\frac{\pi}{\beta}\,G(s,s'),
\end{equation}
where $\langle ...\rangle_{S}$ is the functional average with the
action $S[\Phi]$. For example, the two-level spectral correlation
functions are equal to:
\begin{eqnarray}
\label{R-G}
R_{\infty}(s,s')=1+\frac{2}{\pi\beta}\,\partial_{s}\partial_{s'}\,G(s,s')+\frac{2}{\beta\,(2\pi^{2})^
{\frac{2}{\beta}}}\cos(2\pi \,s)\,e^{8\pi\beta^{-1}\,G(s,s')}.
\end{eqnarray}
For $T=0$ the Green's function is
$G(s-s')=-\frac{1}{4\pi}\,\ln(s-s')^{2}$ at $|s-s'|\gg 1$ and one
reproduces the asymptotic form of the Wigner-Dyson correlations.

In order to reproduce by the same token the corresponding
correlation functions \cite{KTs} for the {\it non-invariant} CRMT
Eq.(\ref{cRMT}) one merely substitutes in Eq.(\ref{boson-act}) the
finite $T$ given by Rq.(\ref{T}) and replaces the zero-temperature
Green's function by the  finite-temperature one:
\begin{equation}
\label{GGF} G(s-s')=\frac{1}{2\pi}\ln\left(
\frac{\pi\,T}{\sinh(\pi\,T\,|s-s'|)}\right).
\end{equation}

Thus we conclude that the deformation of the Wigner-Dyson  RMT given
by Eq.(\ref{cRMT}) with large but finite $b$, retains the analogy
with the Calogero-Sutherland model. However, instead of the ground
state, the non-invariant CRMT with $\kappa\ll 1 $ is equivalent to
the Calogero-Sutherland model at a small but finite temperature
Eq.(\ref{T}).

It is remarkable that there exists a deformation of the free-boson
functional Eq.(\ref{boson-act})  capable of reproducing the
two-level correlation function for the {\it invariant RMT with
log-square confinement}, including the ghost correlation dip. All
one has to do for that is to replace the action Eq.(\ref{boson-act})
defined on a cylinder by that defined on a curved space-time with a
horizon:
\begin{equation}
\label{curved} \int_{0}^{\infty}d\tau\int_{\infty}^{+\infty}
ds\,[(\partial_{s}\Phi)^{2}+(\partial_{\tau}\Phi)^{2}]\rightarrow
\int
d^{2}\xi\,\sqrt{-g(\xi)}\,g^{\mu\nu}\,\partial_{\mu}\Phi\,\partial_{\nu}\Phi,
\end{equation}
where $g\equiv \det\,g_{\mu \nu}$ with $g_{\mu\nu}$ being the
metric, i.e. $ds^{2}=g_{\mu\nu}\,d\xi^{\mu}\,d\xi^{\nu}$ with
$x^{1}=x$ and $x^{0}=t=-i\tau$. The main requirement for the metric
is that the transformation of co-ordinates to the frame
$(\bar{x},\bar{t})$ where the metric is flat ("Minkovski space" with
$ds^{2}=d\bar{x}^{2}-d\bar{t}^{2}$), is exponential for large enough
$x$ with a factorized $x$ and $t$ dependence, e.g.
\begin{eqnarray}
\label{transf}
&&\bar{x}=\varphi(x)\,\cosh(t/A),\;\;\;\;\bar{t}=\varphi(x)\,\sinh(t/A),\\
&&\varphi(x)\sim {\rm sgn}(x)\,e^{\frac{|x|}{A}}\;\;\;{\rm
at}\;\;\;|x|\gg A=2\pi\,T.
\end{eqnarray}
In order to represent the invariant RMT with an {\it even}
confinement potential, the function $\varphi(x)$ should be {\it
odd}. This last requirement plus the continuity of the function at
$x=0$ necessarily implies  $g_{00}(x=0)=0$, that is $x=0$ is the
horizon. One example of such a metric is:
\begin{eqnarray} \label{examp-met}
&&\bar{x}=A\,\sinh(x/A)\,\cosh(t/A),\;\;\;\bar{t}=A\,\sinh(x/A)\,\sinh(t/A)\\
&&ds^{2}=\cosh^{2}(x/A)\,dx^{2}-\sinh^{2}(x/A)\,dt^{2}\\
\nonumber &&
\sqrt{-g(\xi)}\,g^{\mu\nu}\,\partial_{\mu}\Phi\,\partial_{\nu}\Phi=\\
&&=|\tanh(x/A)|\,(\partial_{x}\Phi)^{2}+
|\coth(x/A)|\,(\partial_{\tau}\Phi)^{2}.
\end{eqnarray}
It maps the strip $x\geq 0$, $0<\tau< 2\pi\, A$ and the strip $x\leq
0$, $0<\tau<2\pi\,A$ {\it separately} onto the entire plane
$(\bar{x},\bar{\tau})$.

In order to compute the Green's function on the curved space-time
corresponding to Eq.(\ref{transf}) we use the well known formula
\cite{Tsvelik}:
\begin{equation}
\label{Gr-Gr}
G(z,z')=-\frac{1}{2\pi}\,\ln|\bar{z}(z)-\bar{z}(z')|-\frac{1}{4\pi}\,\ln[|\partial_{z}\bar{z}(z)\,\partial_{z'}
\bar{z}(z')|],
\end{equation}
where $\bar{z}=\bar{x}\pm i\bar{\tau}$, $z=x\pm i\tau$ and
$\partial_{z}=\partial_{x}\mp i\partial_{\tau}$ with $\pm={\rm
sgn}(x)$. Then   from Eq.(\ref{transf}) for $x,x'$ sufficiently far
from the origin one easily obtains:
\begin{eqnarray}
\label{GF} G(x,x')&=&-\frac{1}{4\pi}\ln\left[
\frac{(\varphi(x)-\varphi(x'))^{2}}{4|\varphi(x) \,\varphi(x') |}\right] \nonumber\\
&= &-\frac{1}{2\pi}\left\{
\begin{matrix} \ln\left[2A \sinh[(x-x')/2A]\right], &
x\,x'>0\cr \ln\left[2A \cosh[(x+x')/2A]\right], & x\,x'<0\cr
\end{matrix}\right.
\end{eqnarray}
The origin of $\sinh$ and $\cosh$ in Eq.(\ref{GF}) is very much the
same as that in Eqs.(\ref{kappa-A}),(\ref{ghost}). The exponential
transformation of coordinates Eq.(\ref{transf}) plays the same role
as the exponential unfolding Eq.(\ref{unfold-ln}). Finally
substituting Eq.(\ref{GF}) into the expression for the two-level
correlation function Eq.(\ref{R-G}) one reproduces
Eqs.(\ref{kappa-A}),(\ref{ghost}).

Note that the new scale $A=(2\pi\,T)^{-1}$  sets the temperature
scale $T$ given by Eq.(\ref{T}) which has a meaning of the {\it
Hawking temperature} in the black hole analogy. While the finite
temperature arises because of the periodicity of transformations
Eq.(\ref{examp-met}) in the imaginary time $\tau=i\,t$, this
compactification is different from the standard one: each of the
semi-strips $x\geq 0$ and $x\leq 0$ are mapped on the entire plane
$(\bar{x},\bar{\tau})$ independently of each other. This is
essentially the effect of the horizon.

\sect{Conclusion} I would like to close this Chapter by some
concluding remarks. First of all, it is by no means a comprehensive
review but rather some introduction to the subject written by a
physicist motivated by the physics applications and not by the
formal rigor. My goal was to show that the subject is rich and
poorly explored and that the efforts are likely to be rewarded by
non-trivial discoveries. Let me formulate in the end the (highly
subjective) list of open problems as I see them.
\begin{itemize}
\item
{\it further study of the non-invariant Gaussian critical RMT}
\\ Some progress has been made in this direction by development of
the regular expansion in $b\ll 1$, the so called {\it virial
expansion method} \cite{YK2003,KYC2006,YO}, which extends ideas of
Refs.\cite{Lev,ME}. This approach made it possible to compute
analytically the level compressibility up to $b^{2}$ terms and find
an extremely good and simple approximation \cite{YO,KC} to the LDoS
correlation function Eq.(\ref{LDoS}). These works are the basis for
a perturbative proof  of the dynamical scaling
Eq.(\ref{Chalker-anzats})\cite{Chalker} and the duality relation
Eq.(\ref{dual}).
\item
{\it Non-perturbative solution to the non-invariant critical RMT}\\
Some nice relationships such as the duality relation
Eq.(\ref{dual}), raise a question about a possibility of exact
solution to the CRMT. In our opinion this possibility does exist.
\item {\it Invariant-noninvariant correspondence}\\
This is a very interesting issue with lots of applications in case
the origin of this correspondence is understood. It is also
important to invest some efforts to study the issue of the
correspondence between the spectral and eigenvector statistics, in
particular the  possible spontaneous breaking of unitary invariance
and emergence of a preferential basis \cite{KrCan}.
\item {\it Level statistics in weakly non-self-unfolding RMT}\\
This is a broad class of invariant RMT where the Wigner-Dyson
universality is broken. The spectral statistics in such RMT exhibits
unusual features like super-strong level repulsion near the origin.
\item {\it Crystallization of eigenvalues}\\
This phenomenon takes place in the non-perturbative regime
$\kappa>1$ of RMT with log-square confinement and has its
counterpart in the "crystallization" of roots of orthogonal
polynomials \cite{Bogomolny-Pato}. So far it does not have physical
realization, with Calogero-Sutherland model being a clear candidate
\cite{Hald}.
\end{itemize}

\ \\
{\sc Acknowledgements}: I am grateful to my coworkers C.M.Canali,
J.T.Chalker, E.Cuevas, F.Franchini, I.V.Lerner, M.Ndawana,
K.A.Muttalib, A.Ossipov, A.M.Tsvelik and O.Yevtushenko for extremely
insightful and enjoyable collaboration.

\end{document}